\documentclass{ws-p8-50x6-00}

\newcommand{\BEQ}{\begin{equation}}

\newcommand{\EEQ}{\end{equation}}

\newcommand{\BEA}{\begin{eqnarray}}

\newcommand{\EEA}{\end{eqnarray}}

\renewcommand{\d}{{\rm d}}

\renewcommand{\top }{ t^{\prime } }

\newcommand{\W}{\rho}

\begin{document}

\title{From Fermi-Pasta-Ulam Problem to Galaxies: The Quest for Relaxation}

\author{A.E.Allahverdyan$^{1,2}$ and V.G.Gurzadyan$^{2,3}$} 

\address{1. SPhT, CEA Saclay, 91191 Gif-sur-Yvette cedex, France;
2.Yerevan Physics Institute, Yerevan, Armenia; 3.ICRA, Department of Physics, 
University of Rome La Sapienza, Rome, Italy}

\maketitle

\abstracts{
The major problem posed by the Fermi-Pasta-Ulam work was the relaxation
(thermalization) of a many-body system.
We review two approaches to this problem, the ergodic theory
and Langevin (stochastic) equation ones, which have been 
applied for the description of statistical mechanics of stellar systems. 
Among the particular problems that we consider, are 
1-dimensional long-range interacting systems and iterated maps and
3-dimensional N-body gravitating systems.  
}

\section{The Fermi-Pasta-Ulam system: Historical introduction}

The numerical experiment by Fermi, Pasta and Ulam opened several new
fields of research and appeared to be an invaluable source of
inspiration for generations of physicists.

This work is outstanding at least by the following reasons:

1. It represents the first computer study of a nonlinear system;

2. The results contradicted the belief held since Poincare. Fermi had
considered it 'a little discovery' (as quoted by Ulam), thus
immediately evaluating their extraordinary importance;

3. It was one of last Fermi's works, completed after his ultimate
death in 1954;

4. Remained unpublished during a decade; 

5. Coincides in time with Kolmogorov's theorem (1954), 
though FPU and Kolmogorov-Arnold-Moser (KAM) theory 'met' only in
1966. 

6. Inspired the discovery of solitons and numerous other studies;

7. Its results are not fully understood till now
and the FPU model continues its inspiring mission also today, after half a century.

The circumstances around this work are now widely known, mainly due to the reminiscences of 
Fermi's collaborator of Los Alamos period and coauthor of this and other works, 
renown mathematician Stanislav Ulam. We mention some aspects only.

The study was planned in the summer of 1952, during one of Fermi's visits to Los Alamos, the main calculations were carried on in the summer of 1953, and final calculations in early 1955. 
It was documented as Los-Alamos internal report LA-1940 in May 1955 \cite{fermi}.

The calculations were performed on MANIAC, one of the first computers, 
together with the programming assistant John Pasta
and with the help of Mary Tsingou.  

Though Ulam had reported the results in several meetings and had published their short account \cite{ulam1}, the report LA-1940 was fully published only in 1965, in the second volume of collected papers of Fermi \cite{fermi}. After that, Ulam had studied other systems as well, i.e. described by homogeneous quadratic transformations, a class of so-called Binary Reaction Systems in three and four variables
and of cubic transformations in three variables, in collaboration with P.Stein and M.Menzel (Tsingou) \cite{ulam3}.

What was the aim of FPU work?

According to Ulam \cite{ulam1}: 

{\it "The ultimate aim was to discuss problems with more independent variables in the hope of obtaining material which would suggest some general features of behavior of systems with an infinite number of degrees of freedom, with nonlinear interaction terms as they occur between the oscillators in quantum theory or between the degrees of freedom in an electromagnetic field or a meson field. The mathematical possibilities concerning what might be called {\it quasi-states} were discussed with Fermi in this connection".}

FPU system represents a 1-dimensional system of N nonlinearly coupled oscillators given by a Hamiltonian
\begin{equation}
H(p,q) = \sum_{j=1}^{N+1}[\frac{1}{2}p_j^2 + \frac{1}{2} (q_j - q_{j-1})^2 + 
\frac{\alpha}{3}(q_{j} - q_{j-1})^m], 
\end{equation}
at boundary conditions
$$
q_j=q_{j+N}.
$$
The calculations were performed for three type of perturbations, cubic, quadratic and of broken-linear function.

At $m=3$ ($\alpha$-FPU) and $4$ ($\beta$-FPU) and $N=16, 32, 64$ with zero velocities of particles at $t=0$, the system was run during time scales corresponding up to 80,000 cycles of the linear system.

Instead of expected thermalization, i.e. a trend towards equipartition over modes, a regular behavior has been observed, namely, "mode 1 comes back to within one percent of its intimal value so that the system seems to be almost periodic" \cite{fermi}.

Ulam recounts \cite{ulambook}:

{\it The results were entirely different qualitatively from what even Fermi, with his great
knowledge of wave motions, had expected.}

Indeed, it was contradicting the widely held views based on the Poincare theorem, 
that such perturbed systems have to be chaotic.

By dramatic coincidence, at the same time that belief was shaken by a theorem published without a proof by Kolmogorov \cite{kolm} in Moscow in 1954. Its proof  was given almost a decade later, by Arnold and Moser in early 1960s, and Kolmogorov-Arnold-Moser (KAM) theory had to become an  achievement of fundamental importance.

Fermi was well familiar with Poincare theorem.
Already in 1923 he had applied Poincare's method to prove that 
a system which has no other integrals than the energy one, has to be ergodic on the energy hypersurface \cite{fermi23}.

It might be, though Ulam mentioned about this indirectly \cite{fermi}, 
that one of Fermi's motivations for FPU study was to check his theorem via an experiment.   
Does the fact that Fermi immediately accepted the paradoxical result of the experiment, indicate his anticipation of new fundamental insights? What would be his reaction, if informed about Kolmogorov's theorem? 

Unfortunately these questions remained unanswered. Fermi died, even before FPU was completed, without delivering his scheduled Gibbs Lecture for 1955 in American Mathematical Society.  

Kolmogorov's theorem needed around 15 years to reach the wider physical audience in west.

Thus, FPU emerged both at the time of major breakthrough in the theory of perturbed nonlinear dynamical systems
and at the moment of appearance of the first electronic computers. 

Fermi's intuition had clearly revealed both these aspects already at that time, as
Ulam refers \cite{fermi} to Fermi's opinion on the importance of the {\it 'understanding of nonlinear systems'} for the future fundamental theories, and the {\it 'potentialities of the electronic computing machines'} and even mentions about Fermi's learning of the actual coding (programming) during one summer.   

The forthcoming decades indeed were marked by spectacular achievements both in the studies of nonlinear systems 
and in the development of the computer technique.

In 1958 Kolmogorov introduced the metric entropy, 
now known as Kolmogorov-Sinai (KS) entropy, as an invariant of the transformation, In 1959 he
introduced K-systems, now known as Kolmogorov systems\footnote{Kolmogorov's original notation for K-systems implied 'quasi-regular systems'.}.

Introducing the entropy as a metric invariant, Kolmogorov moved to probabilistic and geometric methods from the functional approach dominating since von Neumann and Hopf.

His works together with those of Sinai and Anosov in 1960s are considered to lead to the modern ergodic theory of smooth systems\cite{katok}.

FPU work stimulated important studies, already before the publication in Fermi's volumes.  We mention only two.
Tuck and Menzel (Tsingou) in 1961 observed a superperiod at more extensive studies of the same system. Their results remained unpublished until 1972, though Ulam mentioned them when commenting Fermi's volume in 1965 \cite{fermi}.
Still unpublished, FPU inspired the discovery of solitons by Zabusky and Kruskal \cite{zab} in 1965, "as solitary wave pulses which pass through each other and do not sum
in a usual linear-wave superposition."

In forthcoming years FPU continued and still is continuing to serve as a reference point for numerous
studies (for reviews and references see\cite{fordj}-\cite{ruffo}). 
Already the fact, that the decade delay of its publication, which might be fatal for any other scientific result, had not affected its influence, indicates its extraordinary power.

In the present review, on the example of FPU, we will discuss the problem of thermalization of nonlinear
systems associated with an important astrophysical problem, the relaxation of N-body gravitating systems. The structure of the paper is as follows.

First, we briefly discuss the association of FPU to 
the Poincare theorem and KAM theory.
Then, we review the approach to the thermalization based on the
Langevin equation, as of the simplest and most transparent version 
of the Brownian dynamics.
Some results on the long-range interacting systems and iterated maps
are presented thereafter and finally, we discuss the geometrical approach in
stellar dynamics. Each topic is rather broad, and therefore we confine
ourselves with brief accounts, without going into the details of applications.

\section{Relaxation in Statistical Mechanics}

Thus, the FPU puzzle concerned the thermalization, or relaxation towards equilibrium, 
one of the oldest problems in statistical physics. 

On the one hand, the relaxation verifies the {\it
equilibrium} statistical mechanics, since provides the
basic equilibrium distribution functions | microcanonical, canonical
and grand canonical | as the result of a dynamical
relaxation process. On the other hand, the problem of thermalization
constitutes the very subject of non-equilibrium statistical
mechanics \cite{klim}$^{,}$\cite{balian}, since it is supposed to 
describe scenarios of relaxation
and to give information on relaxation times scales. 

Historically, the first step in this field was made by Boltzmann.
His kinetic equation for weakly non-ideal dilute gases was the first
physical example which {\it derived} thermalization starting from the microscopics
dynamics\cite{klim}$^{,}$\cite{balian}. In Boltzmann's setup one considers a closed macroscopic
system, and makes certain sensible {\it kinetic assumptions} on its dynamics which
result in the kinetic equation. This is sufficient to derive the H-theorem,
which states a monotonous increase of entropy, and as the main result
one gets relaxation of the macroscopic system towards equilibrium
(thermalization). 

Unfortunately, this approach is not always easy to
motivate and implement. In particular, the kinetic assumptions
become mathematically
rigorous only in rather artificial limits. For instance, it is known
that a self-consistent description of thermalization for essentially
non-ideal gases within Boltzmann's approach meet serious problems \cite{klim}.

The example of Fermi-Pasta-Ulam comes to verify that not all realistic 
systems need to satisfy conditions of Boltzmann's kinetic equation. 

Another approach to the thermalization problem was started by
Einstein and Langevin \cite{landau,klim,balian}. This is a drastic change in
philosophy as compared with the Boltzmann's approach, since one does not
look for thermalization of the whole macroscopic system, but rather
investigates its subsystem. This approach originated from the
problem of Brownian motion, where one follows to the motion of one
heavy particle immersed in a thermal bath. 

In contrast to the Boltzmann's one, in this approach the 
thermalization problem appears to be more transparent and more widely
applicable, since after
certain reasonable and rather universal assumptions on the structure of
the bath, one derives thermalization of the Brownian particle via an
exact dynamical analysis \cite{klim,balian,weiss,risken}. 

As a consequence, one gets relaxation towards Gibbs distribution. One
of the essential advantages of this approach is that the extension to
quantum-mechanical systems appears to be straightforward \cite{weiss}.

The third approach is provided by ergodic theory which 
classifies dynamical systems by so called statistical
properties, including those of tending to the equilibrium state.

In particular, the property of mixing is defined via decay of
time correlation functions, as necessary condition for relaxation,
while the ergodicity alone is not sufficient for that.

Though this approach is known by difficulties of
checking the sufficiency conditions for realistic physical problems, there
are methods, among them are the
geometrical ones, which enable to overpass those difficulties.

The geometric methods of ergodic theory
have been advanced by Kolmogorov, Sinai, Anosov and others 
(see \cite{anosov,arnoldbook}), while
in physical problems they were firstly applied by Krylov \cite{krylov}
based on the early results of Hedlund and Hopf. 

In stellar dynamics, i.e. for the problem of N-body gravitating systems
interacting with Newtonian potential, the Langevin equation
approach has been used long time ago by Chandrasekhar and von Neumann \cite{chand43,chandvon}.
The geometrical methods of ergodic theory firstly have been applied in stellar dynamics 
in \cite{GS1,GS2,GK86,GK88,GK88a,GKM} (and their other papers quoted therein).

\section{Poincare theorem, KAM and FPU}

First, let us recall that N-dimensional system is considered as integrable
if its first integrals $I_1,...,I_N$ in involution are known,
i.e. their Poisson brackets are zero. 
Then, as follows from the Liouville theorem if the set of 
levels 
$$
M_I=\{I_j(x)=I^0_j, j=1,...,N\}
$$
is compact and connected, then it is diffeomorphic to N-dimensional torus
$$
T^N=\{(\theta_1,...,\theta_N), modd 2\pi\}, 
$$
and the Hamiltonian system performs a conditional-periodic motion on $M_I$.
Poincare theorem \cite{poin} states that {\it for a system with perturbed Hamiltonian
\begin{equation}
H(I, \varphi, \epsilon) = H_0(I) + \epsilon H_1(I, \varphi, \epsilon),
\end{equation}
where $I, \varphi$ are action-angle coordinates, at small $\epsilon > 0$ 
no other integral exists besides the one of energy $H=const$, if
$H_0$ fulfills the nondegeneracy condition,
\begin{equation}
det |\partial \omega /\partial I|\not= 0,
\end{equation}
i.e. the functional independence of the
frequencies $\omega=\partial H_0/\partial I$ of the torus over which the conditional-periodic winding is performed.}

Though this theorem does not specify the behavior of the trajectories of the system
on the energy hypersurface, up to 1950s it was widely believed that
such perturbed systems have to be chaotic. 

Kolmogorov's theorem\cite{kolm} of 1954, now the main theorem of KAM theory,  
came to contradict to that belief, namely, 
formulating the conditions when the perturbed Hamiltonian systems can remain stable. 

It states:

{\it If the system (2) satisfies the nondegeneracy condition (3) and
$H_1$ is an analytic function, then 
at enough small $\epsilon > 0$ most of non-resonant tori, i.e.
tori with rationally independent frequencies satisfying the condition
\begin{equation}
\sum n\omega_k\not=0,
\end{equation}
do not disappear and the measure of the complement
of their union set $\mu(M)\rightarrow 0$ at $\epsilon\rightarrow 0$.}

The theorem  says nothing about the limiting value of the perturbation $\epsilon$. Later numerical studies showed that, for certain systems
that theorem can remain valid even at not small perturbations. Also, the behavior in two-dimensional Hamiltonian systems when the torus separate the regions of three-dimensional system and those of higher dimensions have to be different, when the phase trajectories of the system can wander to other regions due to Arnold diffusion.

Izrailev and Chirikov in 1966 \cite{chirikov}, were the first to match KAM and FPU, via the
formulated resonance overlap (Chirikov) criterion. Though even earlier, in 1963,
Ford and Waters quoted Kolmogorov's address at the International Congress of Mathematicians of 1957: 

{\it "Kolmogorov claims to have shown that nonlinear systems, more general than the systems considered here, are not ergodic"}, the importance of that theorem was not been recognized and the works of Arnold and Moser were unknown.

Now, after numerous studies it is clear that FPU does not satisfy the conditions of the KAM theorem,
though its behavior is close to that of integrable systems and to KAM conditions.
KAM-FPU association is still an open issue.

FPU does not satisfy Poincare theorem conditions either, and cannot be directly traced
via Poincare recurrence time, as was attempted by some authors (see refs in \cite{fordj}). 

The rigorous understanding of the thermalization remains the challenge of FPU.

\section{The Langevin equation}

Let us first consider the derivation of Langevin's equation within the
Brownian motion (system-bath) approach. Instead of describing this
derivation in its full generality, we will consider the simplest
situation as to provide all essential mathematical details, and to
make clear all necessary physical assumptions.

The Langevin equation
is derived from the exact Hamiltonian description of a subsystem
(Brownian particle) and a thermal bath, by tracing out the degrees 
of freedom of the bath. The influence of the particle on the bath is 
assumed to be sufficiently small. Thus, only the linear modes of the 
bath are excited, and the interaction of the particle
with the bath is assumed to be linear. To be as pedagogic as possible,
we first take a definite model for the bath, namely a collection of 
harmonic oscillators.

For the total Hamiltonian we thus assume~\cite{weiss}
\begin{eqnarray}
\label{hamiltonian}
H_{tot}=H+H_B+H_I=\frac{p^2}{2m}~+&&V(x)+\sum_{i}\left [
\frac{p_i^2}{2m_i}+\frac{m_i\omega_i^2}{2}x_i^2\right]\nonumber\\
&&+\sum_i\left[-c_ix_ix+\frac{c_i^2}{2m_i\omega_i^2}\,x^2\right ],
\end{eqnarray}
where $H$, $H_B$ are the Hamiltonians of the particle and the bath,  
and $H_I$ is the interaction Hamiltonian.
$p$, $p_i$, $x$, $x_i$ are the momenta and coordinate  
of the Brownian particle and the linear modes of the bath. 
$c_i$ are coupling constants with the bath; their concrete expressions
will be given later on.
$V(x)$ is the confining potential of the particle, 
and $m$ and $m_i$ are the corresponding masses. 

The Brownian particle was taken to be one-dimensional. This is by no
means a restriction of generality, because the generalization towards
three-dimensional particle, as well as $N$ interacting Brownian
particles, is straightforward.

Notice that the $x_i$-terms form a complete square,
since it includes a self-interaction term proportional to $x^2$.
This guarantees that 
the total Hamiltonian $H_{tot}$ will be positive definite.

At the initial time $t=0$ the bath and the particle are decoupled. 
The distribution function of the particle is not specified. 

The bath, on the other hand, is assumed to be in equilibrium at 
temperature $T=1/\beta$ (we set Boltzmann's constant 
$k_B=1$ in the body of this work), 
and described by its Gibbs distribution.

This means 
\begin{equation}
\label{b33} 
\rho _{SB}(0)=\rho (0)\rho _{B}(0)=
\rho _{S}(0)\frac{\exp [-\beta H_B(0)]}{Z_B},
\end{equation}
where $\rho (0)$, $\rho _{B}(0)$ are the initial distribution
functions of the particle and the bath, and 
$$Z_B=\int\prod_i\d p_i\,\d x_i\,~\exp[-\beta H_B]$$ is the
partition sum of bath.

The Hamilton equations of motion for the bath modes read
\begin{eqnarray}
\label{vavilon1}
&& \dot{x}_i= \frac{1}{m_i}p_i,
\\
&& \dot{p}_i= - x_im_i\omega _i^2 +c_i x
\label{vavilon2}
\end{eqnarray}
These equations are solved readily:
\BEA \label{xit=}
x_i(t)&=&x_{i}(0)\cos\omega_i t+\frac{p_{i}(0)}{m_i\omega_i}\sin\omega_it +
\frac{c_i}{m_i\omega_i}\int_0^t\d s\,\sin\omega_i(t-s)x(s)\\
\label{pit=}
p_i(t)&=&-m_i\omega_i x_{i}(0)\sin\omega_i t+
p_{i}(0)\cos \omega_it +
{c_i}\int_0^t\d s\,\cos\omega_i(t-s)x(s)
\EEA
Thus, the Heisenberg equations of motion for the Brownian particle read
\begin{eqnarray}
\label{vavilon6}
&& \dot{x}= \frac{1}{m}p,
\\
&& \dot{p}= - V'(x) +\sum _ic_i x_i - x\sum_i\frac{c^2_i}{m_i\omega ^2_i}
\label{vavilon7}
\end{eqnarray}
Combined with Eq.~(\ref{xit=}) the last equation becomes
\BEA\label{pdot=}
\dot{p}= - V'(x)  - x(0)\gamma (t)-
\int _{0}^{t}\d s \gamma (t-s)\dot{x}(s)+\eta (t),
\EEA
where
\BEA \label{etat=}\eta(t)=&&\sum_i
c_i[x_{i}(0)\cos\omega_it+\frac{p_{i}(0)}{m_i\omega_i}
\sin\omega_i t]\nonumber\\=
&&\sum_i\sqrt{\frac{\hbar c^2_i}{2 m_i\omega _i}}[a^{\dagger}_i(0)e^{i\omega _it}
+a_i(0)e^{-i\omega _it}],
\EEA
\BEQ \label{hh}\gamma (t)=\sum_i\frac{c^2_i}{m_i\omega ^2_i}
\cos (\omega_i t),
\EEQ
are the noise related to the {\it unperturbed} bath,
and the friction kernel, respectively. Notice that in this exact derivation
the back-reaction of the bath on the particle has been taken into account. 

It is described by the integrals in eqs. (\ref{xit=}), (\ref{pit=}), 
and brings the damping terms $x(t)\gamma(0)-x(0)\gamma (t)-
\int _{0}^{t}\d s \gamma (t-s)\dot{x}(s)$ in eq. (\ref{vavilon7})
to yield (\ref{pdot=}).

\subsection{Drude-Ullersma spectrum}

For some, but not all, applications it is benefitable to
consider a fully explicit case for the bath.
The bath is assumed to have uniformly spaced modes 
\BEQ \omega_i=i\Delta\qquad i=1,2,3,\cdots \EEQ 
and for the couplings we choose the Drude-Ullersma 
spectrum \cite{weiss,ullersma}
\BEQ \label{ci=}
c_i=\sqrt{\frac{2\gamma m_i\omega_i^2\Delta}{\pi}\,
\frac{\Gamma^2}{\omega _i^2+\Gamma^2}}
\EEQ 
Here $\Gamma$ is the characteristic Debye cutoff frequency of the bath,
and $\gamma$ stands for the coupling constant; it has dimension 
$kg/ s$. 

The thermodynamic limit for the bath is taken  by sending 
$\Delta\to 0$. Notice that this creates an infinite ``Poincare''
timescale $1/\Delta$, implying that in the remaining approach the 
limit of ``large times'' always means the quasi-stationary
non-equilibrium state where time is still much less
than $1/\Delta$. 

In the limit $\Delta\to 0$ each coupling $c_i\sim\sqrt{\Delta}$
is very weak. The fact that the bath has many modes nevertheless
induces its non-trivial influence.

At finite but small $\Delta$ the system has an initial relaxational 
behavior, which at times of order $1/\Delta$ is changed in a 
chaotic behavior.

It is customary to define the spectral density
\BEQ J(\omega)=\frac{\pi}{2}\sum_i \frac{c_i^2}{m_i\omega_i}
\delta(\omega-\omega_i)=
\frac{\gamma \omega\Gamma^2}{\omega^2+\Gamma^2} 
\EEQ
It has the Ohmic behavior $ J\approx\gamma\omega$ for
$\omega\ll\Gamma$,
and $\gamma$ is called the interaction strength or damping constant. 

As $J(\omega)$ is cut off at the ``Debye'' frequency $\Gamma$, 
it is called a quasi-Ohmic spectrum.

It can then be shown that the friction kernel (\ref{hh}) becomes 
\BEQ \label{cisum=}
\gamma (t)=\frac{2\gamma }{\pi}\,
\int_0^\infty \d\omega\,
\frac{\Gamma^2}{\omega^2+\Gamma^2}\,\cos\omega t=
\gamma\Gamma\,e^{-\Gamma \,|\,t|}
\EEQ
It is non-local in time, but on timescales much larger than $1/\Gamma$
it may be replaced by $\gamma\delta_+(t)$.

The resulting Langevin equation reads
\begin{equation}
\label{01}
\dot{p}+\frac{\gamma\Gamma}{m}\int _0^t\d t^{\prime}
e^{-\Gamma(t-t^{\prime})}p(t^{\prime}) +V^{\prime}(x)
=-\gamma\Gamma e^{-\Gamma t}x(0)+\eta (t),
\end{equation}
It follows from (\ref{b33}) that the noise is stationary Gaussian, and has
the following properties:
\begin{equation}
\label{02}
K(t-t')=\langle \eta (t) \eta (t')\rangle =
\frac{2 \gamma T}{\pi}\int_0^{\infty}\d \omega 
\frac{\cos\omega (t-t')}{1+(\omega /\Gamma )^2}=2 \gamma
T\,\frac{\Gamma}{2}e^{-\Gamma |t-t'|},
\end{equation}
where, again, the average is taken over the initial state of the bath 
(over realizations of the noise).

The connection between properties of the noise and the friction kernel
is the consequence of fluctuation-dissipation theorem 
\cite{landau,klim,balian,weiss,risken}.

The classical white noise situation 
$K(t)\to 2\gamma T\delta(t)$ is recovered by
taking the high-$\Gamma$ limit (Ohmic spectrum of
the bath frequencies).

In this Ohmic regime one gets finally for $t>0$:
\begin{equation}
\label{1.2}
\dot{p}+\frac{\gamma}{m}p(t)+ V^{\prime}(x)=\eta (t),\qquad
\langle\eta(t)\eta(t')\rangle=2\gamma T\delta(t-t').
\end{equation}
Finally we wish to mention that there are alternative ways
to derive the Langevin equation \cite{landau,weiss}, since
many of its properties are rigidly determined by general statements
like the fluctuation-dissipation theorem \cite{landau}.

Nevertheless, we choose to focus on concrete models, 
because they show in detail how the Langevin 
equation arises from first principles, and thus are better 
suited for pedagogical purposes.

\subsection{Fokker-Planck equation}

Once we are interested by one-time quantities, the state of the
Brownian particle is described by the distribution function $\W(p,x,t)$:
\begin{equation}
\label{wigner}
\W(p,x,t)=
\int \d p_0~\d x_0 \W(p_0,x_0,0)
\langle \delta (p(t)-p)\delta (x(t)-x)\rangle,
\end{equation} 
where the average is taken with respect to the noise,
$\W(p,x,t)$ and $\W(p_0,x_0,0)$ are final and initial 
distribution functions,
while $p(t)$, $x(t)$ are the solutions of (\ref{1.2}) 
for the corresponding initial conditions, and 
for a particular realization of the Gaussian noise.

The Gaussian noise is distributed according to the functional
\begin{equation}
\label{fun}
{\cal P}[\eta ]\sim 
\exp-\frac{1}{2}\int\d t\d s\, \eta(t)K^{-1}(t-s)\eta(s).
\end{equation}
It will be clear that the distribution function $W(y_1,y_2,t)$,
and the propagator 
$\langle \delta (p(t)-y_1)\delta (x(t)-y_2)\rangle $ satisfy the same
differential equation although with different initial conditions.

Therefore, we will use them interchangeably, making additional specifications
only when it will be necessary.

Differentiating $\W(y_1,y_2,t)$ we get 
\begin{equation}
\label{11}
\frac{\partial \W(y_1,y_2,t)}{\partial t} =
-\sum_{k=1}^2\frac{\partial (v_k \W)}{\partial y_k} - \frac{\partial }
{\partial y_1}
\langle \delta (p(t)-y_1)\delta (x(t)-y_2)\eta (t)\rangle,
\end{equation}
where 
\begin{equation}
\label{11.1}
v_1=-\frac{\gamma}{m}p(t)-V^{\prime}(x),\ \
v_2=\frac{p}{m}.
\end{equation}
are the damped Newtonian acceleration and the velocity, respectively.

The following two relations can be verified 
\begin{equation}
\label{12}
\eta (t){\cal P}[\eta ]
=-\int \d \top K(\top -t)\frac{\delta {\cal P}[\eta ]}{\delta \eta (\top )},
\end{equation}
\begin{equation}
\label{13}
\left [\begin{array}{r}
\delta p(t)/\delta \eta (t^{\prime})\\
\delta x(t)/\delta \eta (t^{\prime})
\end{array}\right ]
=\theta (t-t^{\prime})
\left \{ \exp 
\int^t_{t^{\prime}}\d u 
A(x(u))\right \} _{+}
\left [\begin{array}{r}
1 \\
0
\end{array}\right ], 
\end{equation}
\begin{equation}
\label{13a}
A(x)=\left (\begin{array}{rr}
-\gamma /m & -V^{\prime\prime}(x)\\
1/m & 0
\end{array}\right ),
\end{equation}
where $\{ ... \}_{+}$ means the chronological ordering (time-ordering).

The first equality holds since ${\cal P}[\eta]$ is Gaussian; 
the second follows directly from Eq. (\ref{1.2}) because of the relations
\begin{eqnarray}
\label{shut}
&& \frac{\d }{\d t} \frac{\delta p(t) }{\delta \eta (\top )}=
-\frac{\gamma }{m}\frac{\delta p(t) }{\delta \eta (\top )}-V^{\prime\prime}(x)
\frac{\delta x(t) }{\delta \eta (\top )}+\delta (t-\top ), \\
&&\frac{\d }{\d t} \frac{\delta x(t) }{\delta \eta (\top )}=
\frac{1 }{m}\frac{\delta p(t) }{\delta \eta (\top )}
\end{eqnarray}
Using Eqs.~(\ref{11}, \ref{12}, \ref{13}, 
\ref{13a}) we obtain
\begin{equation}
\label{14}
\frac{\partial \W(y_1,y_2,t)}{\partial t}=
\sum_{k=1}^2 \frac{\partial }{\partial y_k}\left \{- v_k \W
+\frac{\partial }{\partial y_1} 
\langle \delta (p(t)-y_1)\delta (x(t)-y_2)
\Phi _{k1} (\{ x \},t)
\rangle \right \},
\end{equation}
where $\Phi$ is the $2\times 2$ matrix
\begin{equation}
\label{144}
\Phi(\{x\},t)=\int_0^t \d \top K(\top)
\left \{ \exp  \int^t_{t-\top }\d u A(x(u)) \right \}_{+} 
\end{equation}
and $\Phi_{k1}$ is the corresponding matrix element
$K(\top )= 2\gamma T\delta(\top )$, implies
\BEA
\Phi _{k1}=\gamma T\delta _{k1}.
\EEA
The final result is that we obtain a diffusion-type equation 
(Kramers-Klein equation) for $\W$ itself \cite{klim,balian,risken}:
\begin{equation}
\label{ko1}
\frac{\partial \W(p,x,t)}{\partial t}=
-\frac{p}{m}\frac{\partial \W}{\partial x}+\frac{\partial}{\partial p}
\left ([\frac{\gamma}{m}p+V^{\prime}(x)]\W \right )+
\gamma T\frac{\partial ^2\W}{\partial p^2}.
\end{equation}

\subsection{Thermalization and H-theorem}

It is now possible to provide a general argument that Eq.~(\ref{ko1})
ensures thermalization. First one notices that its stationary
solution, obtained from (\ref{ko1}) by setting $\partial_t\W=0$,
is given by the standard Gibbs distribution:
\BEA
\label{st}
\W_{st} =\frac{1}{Z}\,\exp \left[-\beta \left(\frac{p^2}{2m}+V(x)\right)\right],
\EEA
where $Z$ is the statistical sum. Second, one notices that the
H-function \cite{berg,sh,risken}
\BEA
\label{h}
H(t)=\int\d p\,\d x\,\W(p,x,t)\ln\frac{\W(p,x,t)}{\W_{st}(p,x,t)},
\EEA
where $\W(t)$ is the actual time-dependent solution, monotonously
decreases towards zero, which is its equilibrium value $\W=\W_{st}$.

Indeed, differentiating $H(t)$, using (\ref{ko1}) and requiring that
both $\W$ and $\W_{st}$ are zero at infinity, one gets:
\BEA
\label{hdot}
\dot{H}(t)=-\gamma T\int\d p\,\d x\,\W(p,x,t)\left[
\frac{\partial}{\partial p}\,\ln\frac{\W(p,x,t)}{\W_{st}(p,x,t)}
\right]^2<0.
\EEA
Thus, all solutions of (\ref{ko1}) relax with time towards the
stationary Gibbs distribution (\ref{st}). Eq.~(\ref{hdot}) provides an
information on the global relaxation time; it is proportional to
the damping constant $\gamma$ and temperature $T$. Notice that the

H-function provides an information-theoretic distance between the
distribution functions $\W_{st}$ and $\W$ \cite{berg,sh}.

%\subsection{Discussions}
Thus, the open system approach provides a rigorous justification of
thermalization. Let us outline the main assumptions of the
present setup.

1) The Brownian particle was assumed to be in contact
with a macroscopic thermal bath at equilibrium. 

2) The influence of the
particle to the bath was assumed to be weak enough, so that the
non-linear modes of the bath are not excited.

3) The choice of initial
conditions (\ref{b33}) was special, and moreover, it has to be special. One can
motivate that for any macroscopic system there are initial states
which do display anti-relaxational behavior \cite{balian}. For
sufficiently simple cases one can show that those states are
exceptional \cite{balian}.

4) The thermalization was obtained via a rigorous approach; no {\it a
priori} assumptions on ergodicity and mixing were used. As the main result
Eq.~(\ref{hdot}) provides a general expression for the relaxation time.

\section{Long-Range Interacting Systems}

The problem of understanding of the observed thermalization in FPU system thus includes the explanation of the existence of large relaxation times $\tau_R$ and their dependence on the number of particles $N$ and the energy density $\epsilon=E/N$ 
(see \cite{ruffo} and references therein).

We will mention some of the remarkable results obtained in the recent period. The following dependence was found\cite{deluca} 
\begin{equation}
\tau_R \propto \frac{\sqrt N}{\epsilon},
\end{equation}
for finite $N$ and fixed wave number interval for the $\beta$-FPU model, i.e. the divergence of the equipartition time scale with increase of $N$. On the other hand, in the thermodynamical limit the dependence on $N$ disappears \cite{cas}
\begin{equation}
\tau_R \propto \epsilon^{-3}.
\end{equation}
The conclusion of these studies is that, "the process of relaxation to equipartition in the FPU model is not regulated by the microscopic chaotic instability but by the typical time in which an orbit diffuses in phase space, which is determined by the interaction among the phonons" \cite{ruffo}.

While moving from FPU to 1 and 2-dimensional models with
long-range interaction, to mimic the gravitational
systems, the problem becomes more complex since acquires new 
thermodynamics, negative specific heat, etc.

Indeed, different results have been obtained at the analysis 
of the chaotic properties on the following 1-dimensional
\begin{equation}
H=\sum_{i=1}^{N} \frac{p^2}{2m} + \frac{m^2}{2N}\sum_{i,j=1}{N}(1-\cos(\theta_i-\theta_j))
\end{equation}
and 2-dimensional systems
\begin{eqnarray}
H&&=\sum_{i=1}^{N} \frac{p_{i,x}^2+p_{i,y}^2}{2m} \\ 
&&+ \frac{1}{2N}\sum_{i,j=1}{N}[3-\cos(x_i-x_j)-\cos(y_i-y_j)-
\cos(x_i-x_j)\cos(y_i-y_j)],\nonumber
\end{eqnarray}
and, particularly of their equipartition time scales. As a result, 
"few scaling laws have been found to be universal for all these
models. In the gaseous phase the maximal Lyapunov exponent vanishes as 
\begin{equation}
N^{-1/3} 
\end{equation}
in analogy found in\cite{GS2} for some specific gravitational systems" \cite{antoni}.

In a similar model, the Konishi-Kaneko iterated map \cite{konishi}
\begin{equation}
\label{R1}
p^{n+1}_i=p^n_i + k\sum_{j=1}^{N}\sin 2\pi(x^n_j-x^n_i),
\end{equation}
\begin{equation}
\label{R2} 
x^{n+1}_i=x^n_i+p^{n+1}_i;  (mod1) .
\end{equation} 
which can be represented as a system with Hamiltonian 
\begin{equation}
H=\sum_{i=1}{N}[\frac{p_i^2}{2} + \frac{k}{4\pi}\sum_{i\not=j}{N}cos(2\pi(x_i-x_j))]
\end{equation}
the appearance of cluster-type configurations, of power law correlations \cite{koyama},
and Feigenbaum period-doubling bifurcations have been observed \cite{gmo}.
Particularly, the bifurcation scale $\delta=8.72...$ can be estimated at the following conditions for
the bifurcation points 
$$
\sum_{j}^{N} |x_{j}^{n+1} - x_{j}^n| < \epsilon, (2^1=2),
$$ 
$$
\sum_{j}^{N} |x_{j}^{n+2} - x_j^n| < \epsilon, \sum_{j}^{N} |x_j^{n+3} - 
x_j^{n+1}| < \epsilon, (2^2=4),
$$
for each $k_n, n=1,2,...$, respectively, where ${\epsilon}$ is the accuracy of the obtained values of $k_n$. 

The use of iterated maps can enable one to avoid
the principal difficulties associated with $N$-body systems -- the 
non-compactness of the phase space and singularity of Newtonian interaction -- which are not always 
crucial at the study of a particular property of a given stellar system. 
Also, as we saw for Konishi-Kaneko map, the iterated maps can be informative in revealing the
mechanisms of developing of chaos. 

\section{Gravitating Systems}

While moving to the gravitating systems first one must recall that, the Henon-Heiles \cite{henon} system being one the first demonstrations of chaos at numerical experiments, was motivated by the study of the motion in the axisymmetric Galactic potential.

Since then, developments in nonlinear dynamics had various impacts on the problem of N-body gravitating systems (for reviews see the volumes \cite{GP}, \cite{GR}). 

The chaotic dynamics had offered principally new insight on the properties and the evolution of the Solar system.
In the works of Laskar \cite{laskar}, Tremaine\cite{trem}, Wisdom\cite{wis} and their collaborators it was shown that the inner planets can possess chaotic properties influenced by the perturbations of the large planets. Particularly, Mercury and Venus undergo chaotic variations of the eccentricity of their orbit, which can lead to overlapping of orbits with subsequent drastic change of the orbit of Mercury, up to its escape from the Solar system.

By means of the frequency map analysis technique developed by Laskar, the stabilizing role of the 
Moon in the chaotic variations of the obliquity of the Earth has been shown with direct consequences for the climate and hence for the evolution of life on the Earth. Chaotic variations of the obliquity were shown to be important for Mars.

The idea of the frequency map technique \cite{laskar} is in the search of quasiperiodic approximations of the solutions of the perturbation problem by means of finite number of terms
\begin{equation}
x_j(t)=x^0_j + \sum_{k=1}^{N}a (m_k; \nu) \exp[i<m_k,\nu >t]
\end{equation}
where $\nu_j$ are the frequency vectors. Note, that the system is close to an integrable one, and possess KAM (invariant) tori in the phase space.

So, proper theoretical treatment on the perturbed systems together with advanced computations had led to such basic results.

While profound role of chaotic effects in the evolution of Solar system is established, their role in stellar dynamics is often remained underestimated\footnote{See e.g. \cite{heggie}, where the authors needed half a page to transfer from linear to nonlinear effects in describing the relaxation of stellar systems.} even though the latter ones are not close to integrability and KAM conditions, and one might expect that chaos has to be of more importance for them than for planetary systems. 

The problem of relaxation of stellar systems can illustrate the situation.

The two-body relaxation formula derived many decades ago via the linear sum of contributions of several two-body interactions, is still the basic reference for stellar system, even though it predicts time scales exceeding the age of such well-relaxed systems as the elliptical galaxies. 

A priori it is not excluded that the two-body relaxation time scale can be the solution of the nonlinear
problem even when the perturbations of other particles are included, however this does not follow from the 'linear' derivation borrowed from plasma physics, involving the Debye screening and the Coulomb logarithm, i.e. ignoring the long range interaction of gravitating particles. 

While even for 1-dimensional systems the relaxation is essentially different for short and long range interactions. 

Geometrical methods appear as a useful tool for the study of nonlinear effects of gravitating systems.
In \cite{GS1},\cite{GS2} the K-mixing and exponential tending to an equilibrium state (quasi-equilibrium) was shown analytically for perturbed spherical N-body systems.

That was reached via the proof that while the two-dimensional curvature of the configuration space for N-body systems is sign-indefinite, for spherical systems in large N limit it is strongly negative.

The formula derived for the relaxation time scale for the parameters of real stellar systems differs from the two-body relaxation formula, and by now is supported by (a) alternative theoretical derivation as mentioned above\cite{antoni}; (b) numerical simulations and (c) observational data of star clusters (see \cite{elzant} \cite{GR} and references therein).

The importance of chaotic effects in relaxation of stellar systems was shown by 
Pfenniger\cite{pfen} using the Lyapunov exponent technique. That study is especially remarkable given the limited applicability of Lyapunov exponents for many-dimensional systems\cite{GK94}.

Back in 1964 at a remarkable computer study of his time Miller
\cite{miller} had observed exponential growth of errors of the 
coordinates and velocities, $\Delta = \sum(\delta x^2+\delta v^2)$ 
for systems with up to 32 gravitating particles. 

The result of Miller remained not well understood for quite a long
time. Recently, in some papers (e.g.\cite{vallur} and refs therein) 
it is identified with the exponential instability (K-mixing) shown 
in \cite{GS1,GS2}, even though their exponential time-scales are 
quite different. Since such opinion is moving from one paper to
another, we will consider this issue in more details.

As follows from the Maupertuis principle, only trajectories in the 
region of the configuration space $M= [E-U(r) > 0]$ defined by the 
Riemannian metric\cite{arnoldbook},\cite{anosov}
\begin{equation}
ds^2=(E-U(r_{1,1}..., r_{N,3})\sum_{a=1}^{N}\sum_{i=1}^{3}dr^2_{a,i}
\end{equation}     
are corresponding to the N-body Hamiltonian system 
\begin{equation}
H(p,r)=\sum_{a=1}^{N}\sum_{i=1}^{3}\frac{p^2_{a,i}}{2m_a} + U(r),
\end{equation}
where
\begin{equation}
U(r)=-\sum_{a<b}\frac{Gm_am_b)}{|r_a-r_b|}.
\end{equation}
The Maupertuis parameterization and hence Riemannian metric guarantee 
the conservation of total energy of the system
$$
H=E=const,
$$
and therefore enables to study the hyperbolicity and mixing of the system with exponentially deviating trajectories.  

The works discussing the exponential instability can be divided into two groups: 

1. Based on Maupertuis parameterization:  by Krylov, Anosov\cite{anosov} and Arnold\cite{arnoldbook}, and in stellar dynamics, \cite{GS1,GS2,GK88};

2. Without using Maupertuis parameterization: numerical experiments 
by Miller\cite{miller}, repeated by Lecar and Standish in 1960s and again in 
1990s in many papers by Kandrup, Goodman, Heggie, Hut and others (see \cite{vallur,heggie} for references).

In the first group of works the exponent is linked with mixing and relaxation via theorems
of ergodic theory. For the second group, no such link exists. The reason is as follows.

Without Maupertuis parameterization, two points on two phase trajectories can deviate while the trajectories themselves may not deviate at all.
Such apparent exponential deviation will be observed for systems of any number of particles, even of very few, and of any configurations, say of disk type. In contrary, the exponential deviation of\cite{GS1,GS2} exists only for spherical systems and only at large N-limit, and not for disk or few-particle gravitating systems.  

Therefore, a statement that, the second group papers also refer to chaos and mixing, is the same as to assign chaos to the two-body system, where the same apparent deviation can be observed.

This is the reason that, only the dynamical (crossing) time scale is observed at such numerical experiments.

Numerical simulations \cite{elzant} taking into account the Maupertuis parameterization are in agreement with the relaxation time scale derived in \cite{GS1,GS2}. 

At the same time, we are not aware of any numerical experiments which confirm the two-body 
relaxation time scale\footnote{Donald Lynden-Bell to whom we consulted (V.G., July,2001), was not aware of such experiments either.}. 

The fact that Miller's experiment does not imply mixing, by no means diminishes the pioneering character of his work. However, 
now the results of theory of dynamical systems provide its
interpretation.

The geometrical approach enabled to arrive also to other general conclusions on stellar systems\cite{GS1}:

{\it elliptical and spiral galaxies should have different origins.}

The spiral galaxies particularly, are more regular than ellipticals, i.e. do not possess the property mixing \cite{GK86}. Due to the existence of fundamental frequencies they can be described by Laskar's frequency map technique \cite{laskarpap}

At Geneva workshop in 1993 one of us (V.G.) presented a list of "10 key problems of stellar dynamics"\cite{10prob}. Now, about 10 years after, one can note that though a considerable work has been done, especially a numerical one due to advances of computer technique \cite{makino}, still there is a long way to go to fully evaluate the role of nonlinear effects for 
stellar systems.

\section{Epilogue}

The Fermi-Pasta-Ulam problem was able to address profound issues of
nonlinear systems, though studied mainly from physical concepts, 
rather than from rigorous mathematical ones.
It still lacks answers to basic questions.   
As mentioned Pesin \cite{pesin}: "It is a very interesting question
the relation between KAM and FPU and whether FPU can serve as an "experiment" for KAM".
One can expect that the activity in FPU problem will shift towards
investigation of its rigorous aspects.

The relaxation problem remains a central issue for
many systems, including open and long-range
interacting ones. In astrophysical context it concerns N-body gravitating systems, 
galaxies and star clusters. We briefly discussed two powerful approaches which have been
applied in stellar dynamics,
the Brownian particle approach and the geometrical methods of ergodic theory.

In stellar dynamics also one has to expect more rigorous studies of nonlinear effects,
including in the performance and interpretation of numerical experiments, a lesson given to us
by Fermi, Pasta and Ulam 50 years ago.

\end{document}